\documentclass[a4paper,11pt,english,headsepline]{article}
\usepackage[pdftex]{graphicx}
\usepackage[T1]{fontenc}
\usepackage{lmodern}
\usepackage{slashed}
\usepackage[utf8]{inputenc}
\usepackage[english]{babel}
\usepackage{microtype}
\usepackage{cite}
\usepackage{csquotes}
\usepackage{amsmath}
\usepackage{amssymb}
\usepackage{amsfonts}
\usepackage[nottoc,notlot,notlof]{tocbibind}
\usepackage{upgreek}
\usepackage{mathtools}
\numberwithin{equation}{section}
\allowdisplaybreaks
\usepackage[all]{xy}
\usepackage{color}
\usepackage{graphicx}
\graphicspath{{images/}}
\usepackage{geometry}
\usepackage[toc,page]{appendix}
\usepackage{hyperref}
\usepackage[normalem]{ulem}

\newcommand{\diff}{\mathrm{d}}

\usepackage{bm}
\usepackage{ragged2e}
\usepackage{appendix}
\usepackage{slashed}
\usepackage{bbold}
\usepackage{cancel}

\usepackage{lmodern}
\usepackage{slashed}
\usepackage{amsmath}
\usepackage{amssymb}
\usepackage{amsfonts}
\usepackage[nottoc,notlot,notlof]{tocbibind}
\usepackage{upgreek}
\usepackage{mathtools}
\numberwithin{equation}{section}
\allowdisplaybreaks
\usepackage[all]{xy}
\usepackage{color}
\usepackage{graphicx}
\graphicspath{{images/}}
\usepackage{geometry}
\usepackage[toc,page]{appendix}
\usepackage{hyperref}
\usepackage[normalem]{ulem}

\usepackage{bm}
\usepackage{ragged2e}
\usepackage{appendix}
\usepackage{slashed}
\usepackage{bbold}
\usepackage{cancel}
\usepackage{cleveref}
\crefformat{section}{\S#2#1#3} 
\crefformat{subsection}{\S#2#1#3}
\crefformat{subsubsection}{\S#2#1#3}

\usepackage{physics}
\usepackage{tikz-cd}
\usepackage{mathrsfs}  
\usepackage{comment}
\usepackage{cancel}
\usepackage{dsfont}
\usepackage{bbold}
\usepackage{dsfont}
\usepackage{empheq}
\usepackage{stackrel}

\usepackage{multirow} 

\usepackage{mathbbol}

\def\noi{\noindent}

\newcommand{\eq}{\begin{equation}}
\newcommand{\eqa}{\begin{eqnarray}}  
\newcommand{\en}{\end{equation}}
\newcommand{\ena}{\end{eqnarray}}
\newcommand{\enn}{\nonumber \end{equation}}

\usepackage{authblk}

\title{{\Huge \bf  Sen's Mechanism }\\ \vspace{.3cm} {\Huge \bf for Self-Dual Super Maxwell theory}}
\author[1]{G. Barbagallo\thanks{g.barbagallo@studenti.unipi.it}}
\author[2,3]{P. A. Grassi\thanks{pietro.grassi@uniupo.it}}
\affil[1]{Dipartimento di Fisica "Enrico Fermi"
	\\Universit\`a di Pisa, Largo Bruno Pontecorvo, 3, 56127 Pisa PI, Italy}
\affil[2]{Dipartimento di Scienze e Innovazione Tecnologica
	\\Universit\`a del Piemonte Orientale, viale T. Michel 11, 15121 Alessandria, Italy}
\affil[3]{INFN, Sezione di 
	Torino, via P. Giuria 1, 10125 Torino, Italy}

\usepackage{color}

\begin{document}

\maketitle
\begin{abstract}
	In several elementary particle scenarios,  self-dual fields emerge as fundamental degrees of freedom. Some examples are the $D=2$ chiral boson, $D=10$ Type IIB supergravity, and $D=6$ chiral tensor multiplet theory. {
	For those models, a novel variational principle has been proposed in the work of Ashoke Sen. The coupling to supergravity of self-dual models in that new framework is rather peculiar to guarantee the decoupling of unphysical degrees of freedom. 
	We generalize this technique to the self-dual super Maxwell gauge theory in  $D=4$ Euclidean spacetime both in the component formalism and the superspace. 
	We use the geometric tools of rheonomy 
	and integral forms since they are very powerful geometrical techniques for the extension to supergravity.} We show the equivalence between the two formulations by choosing a different integral form defined using a Picture Changing Operator. That leads to a meaningful action functional for the variational equations. In addition, we couple the model to a non-dynamical gravitino to extend the analysis slightly beyond the free case. A full-fledged self-dual supergravity analysis will be presented elsewhere. 
\end{abstract}
\newpage
\setcounter{page}{1}


\section{Introduction}\label{intro}

In several interesting supersymmetric models, self-dual fields are essential to guarantee the 
matching between bosonic and fermionic degrees of freedom. For example, the $D=2$ chiral 
boson model, the $D=6$ tensor and supergravity multiplet \cite{cube}, 
the $D=10$ Ramond-Ramond sector of type IIB supergravity suffer from the same problems. More precisely, 
their equations of motion must be supplemented by the self-dual conditions, but both self-duality constraints and 
equations of motion cannot be obtained from an action principle. Indeed, their kinetic terms vanish once the self-duality condition is imposed. 
{{There have been several proposals  to solve this issue, but they have either a limited 
range of applications or require special choices (non-covariant 
representations\cite{Henneaux,9304154,9701008,0605038},
infinite auxiliary fields \cite{Mcclain,Wotzasek,Martin,9603031,Faddeev,9609102,9607070,9610134,9610226}, non-polynomial actions \cite{ 9509052,9611100,9707044,
	9806140,9812170}, one higher dimension \cite{9610234,9912086}).}} To solve this problem, A. Sen, in seminal papers \cite{Sen:2015uaa, Sen:2015nph,  Sen:2019qit}, inspired by string field 
theory methods, provided a suitable action in terms of a set of auxiliary fields, yielding the correct equations of motion, and which do not interact with the rest of the theory. Those fields 
have special transformation properties which do not depend on supergravity fields 
\cite{Sen:2015nph,Sen:2019qit,Andriolo:2020ykk}. Recently, the mechanism has also been applied to M5-brane in the papers \cite{Andriolo:2020ykk}, \cite{Grassi:2022pag}. 
In \cite{Hull:2023dgp}, generalizations of Sen's mechanism for self-dual p-form gauge fields have 
been provided on a general spacetime, thee approaches explore in a more complete way the coupling with supergravity. Recently, in the work 
\cite{Bandos:2020hgy} the competitor PST \cite{9611100} method has been newly used, it would be interesting to see an integral/superspace version of that framework and to compare with Sen's mechanism.     
Very interesting recent works have been put forward in \cite{Evnin:2022kqn, Arvanitakis:2022bnr}, and it would be interesting to compare them with Sen's mechanism approach.

Tackling the problem from another side, it has been noticed \cite{cube} that 
from a purely geometrical perspective in the rheonomy formalism, the superspace equations of 
motion imply the dynamical equations, the parametrization of the curvatures, and the self-duality constraints 
when needed \cite{cube}, \cite{Castellani:1991jf}. Nonetheless, when those equations of motion are projected to space-time, the self-duality constraint 
is lost. This happens in all known cases (for some examples, see the recent \cite{Cremonini:2020skt, Grassi:2022pag} and the considerations therein). Furthermore, the rheonomy superspace equations are not derived from a genuine 
action principle since the action depends on the choice of the embedding of the bosonic submanifold into 
a suitable supermanifold (unless the Lagrangian, seen as a differential form, is closed. For 
further details, see \cite{Castellani:2014goa}). 

It has been observed in \cite{Cremonini:2020skt}, that the rheonomy Lagrangian, dressed with a special operator 
in the superspace, known as Picture Changing Operator (PCO), shares formal structural analogies with string field 
theory action \cite{Sen:2015uaa, Sen:2015nph,  Sen:2019qit,Grassi:2016apf} and this can be used to suitably modify the Lagrangian to reproduce 
the Sen's mechanism in superspace.  Finally, its restriction to spacetime does not lose the self-duality constraints. This has been explicitly verified in \cite{Cremonini:2020skt} for all possible choices
of PCO at the component level and the superspace level. 

We study another interesting case in the present work where the same phenomena appeared. This is the case of self-dual 
super Maxwell theory in Euclidean signature (here we present only the abelian case, the non-abelian will be presented elsewhere) with the 
supersymmetric spectrum made of a positive-helicity gauge boson and a positive-helicity fermion \cite{Gilson:1985uj,Volovich, Berkovits:1997wj, Sokatchev:1995nj, Sezgin:1992ug, Bern:1996ja, Devchand:1992st}
\footnote{{ In the case of $D=(2,2)$ spacetime, there exist an extended literature on self-dual models, see for example  \cite{Ketov:1992ix,Nishino:1992ri, Ketov:1992rh,Gates:1992rg,Siegel:1992xp}. Those models might serve as a starting point for future investigation of the framework presented here. In those papers, 
in particular in  \cite{Ketov:1992ix} the authors implement the self-duality constraints by Lagrangian multipliers adding more degrees of freedom. It would be interesting to see whether a Sen's mechanism could be implemented such that those additional degrees of freedom are decoupled from the rest. In particular, it would be very interesting to develop a superspace version of Sen's mechanism also in $D=(2,2)$ superspace.}
}. In that case, the self-duality can be imposed ab-initio, and the Lagrangian 
for the self-dual gauge boson is a total derivative. At the same time, the Dirac Lagrangian vanishes in the absence of the anti-chiral part of the gaugino.  We provide a component construction for the Lagrangian implementing 
Sen's mechanism, by introducing two auxiliary fields and modifying the Lagrangian,  we obtain the correct equations for the self-dual part of the fields. The auxiliary fields satisfy free equations, and their fluctuations decouple. We show how all steps of Sen's construction can be
performed, and, for an illustrative interacting example, we consider the coupling of the self-dual multiplet with a non-dynamical gravitino \cite{Nishino:2020rga}. In addition, we provide the construction in the geometric framework with suitable PCOs (only two of them have been considered, namely the spacetime PCO and the supersymmetric one). Finally,  we can obtain the component and the superspace action with the correct equations. Together with \cite{Cremonini:2020skt}, this model is an example of a superspace extension of Sen's mechanism. 

In section 2, we recall some basic facts about SYM theory. In section 3
we project the rheonomic action down to spacetime. Section 4 implements Sen's mechanism 
at the spacetime level; section 5 provides its superspace extension.
Finally, in section 6, the analysis with a non-dynamical gravitino is performed.
For detailed calculations, see \cite{Barbagallo}.

\section{Some basic facts about  Super Yang-Mills theory $\mathcal{N}=1, D=4$}
In the usual formulation of super Yang-Mills theory \cite{Muller, Wess:1992cp, Gates:1983nr}, the gauge field is written as a superconnection expanded in the supersymmetric flat basis $ 		V^{\alpha\dot\alpha}=
		\diff x^ {\alpha\dot{\alpha}} -2i(\diff \theta^{\alpha}\bar{\theta}^{\dot{\alpha}} +\diff \bar\theta^{\dot\alpha}  {{\theta}}^{{\alpha}} )$,
		$
			\psi^{\alpha}= \diff \theta^{\alpha} $
		$	
		\bar\psi^{\dot\alpha} = \diff \bar\theta^{\dot\alpha}:
		$
\begin{equation}
	\begin{aligned}
		\mathscr{A}=\mathscr{A}_A e^A =
		\mathscr{A}_{a}V^{a}+\mathscr{A}_{\alpha}\psi^{\alpha}+\mathscr{A}_{\dot{\alpha}}\bar{\psi}^{\dot{\alpha}}
	\end{aligned}
\end{equation}
and its field strength is given by
$$
	\label{Fpsi}\small
	\begin{aligned}
		\mathscr{F}& =\frac12\mathscr{F}_{AB}e^A\wedge e^B=\\&=
		\frac12\bigg(\mathscr{F}_{ab}V^{a}\wedge V^{b}+
		\mathscr{F}_{a\beta }V^{a}\wedge \psi^{\beta }+
		\mathscr{F}_{a\dot{\beta} }V^{a}\wedge \bar{\psi}^{\dot{\beta} }+ 
		\mathscr{F}_{\alpha\beta }\psi^{ {\alpha}}\wedge \psi^{\beta }+
		\mathscr{F}_{\alpha\dot{\beta} }\psi^{ {\alpha}}\wedge \bar{\psi}^{\dot{\beta} }+
		\mathscr{F}_{\dot{\alpha}\dot{\beta} }\bar{\psi}^{ {\dot{\alpha}}}\wedge \bar{\psi}^{\dot{\beta} }\bigg).
	\end{aligned}
$$
Defining $\mathscr{F}$ as usual  $\mathscr{F}\equiv \diff \mathscr{A}+\mathscr{A} \wedge \mathscr{A}$ it satisfies  the  Bianchi identity $\mathscr{D}\mathscr{F}=0$, where $\mathscr{D}$ is the  covariant derivative.  The constraints eliminate the superfluous degrees of freedom:
\begin{equation}
	\label{constarints}
	\mathscr{F}_{\alpha\beta}=0, \qquad \mathscr{F}_{\alpha\dot\beta}=0, \qquad \mathscr{F}_{\dot\alpha\dot\beta}=0.
\end{equation}
That leads to 
\begin{equation}
	\label{FFWW}
		\mathscr{F}= 
		\frac12\bigg(\mathscr{F}^+_{\alpha \beta }  (V^2_+)^{\alpha\beta}
		+\mathscr{F}^-_{\dot\alpha \dot\beta }  (V^2_-)^{\dot\alpha\dot\beta}+
		{i} \overline{W}_{\dot\alpha}(V  \psi)^{\dot\alpha }+
		{i}{W}_{ \alpha} (V\bar\psi)^{\alpha} \bigg)
\end{equation}  
with the following conditions on the gauge   and the gaugino field-strengths $\mathscr{F}^{\pm}$, $W$ and $\overline{W}$
\begin{equation}
	\label{FDWFDW}
	\mathscr{F}^+_{\alpha \beta }=
		-\frac{1}{4}\mathscr{D}_{(\alpha} W_{\beta)}
		,\qquad
		\mathscr{F}^-_{\dot\alpha \dot\beta }=
		-\frac{1}{4}\overline{\mathscr{D}}_{(\dot\alpha}\overline{W}_{\dot\beta)}
\end{equation}
\begin{equation}
	\label{DDD}
	\overline{\mathscr{D}}_{\dot\beta} W_{ \alpha}=0,\qquad
	\mathscr{D}_{\beta}\overline{W}_{\dot\alpha}=0,\qquad
	\overline{\mathscr{D}}_{\dot\alpha}\overline{W}^{\dot\alpha}=\mathscr{D}^{\alpha} W_{ \alpha}.
\end{equation}  
Moreover, if the group is an abelian one, the covariant derivatives are replaced by  $\mathscr{D}_a\rightarrow \partial_a$, $\mathscr D_{\alpha} \rightarrow D_{\alpha}$, $\overline{\mathscr D}_{\dot\alpha} \rightarrow \overline{D}_{\dot\alpha}$ and   the following relations hold   
\begin{equation}
D_{\rho}\mathscr{F}^-_{ \dot\alpha\dot\beta }=-\frac{i}{2}\partial_{\rho(\dot\alpha}\overline W_{\dot\beta)},\qquad
\overline D_{\dot\rho}\mathscr{F}^-_{ \dot\alpha\dot\beta }=\frac{i}{2}\epsilon_{\dot\rho(\dot\alpha}\partial_{\dot\beta)\rho}  W^{ \rho}
\label{DF-}
\end{equation}
\begin{equation}
\overline D_{\dot\rho}\mathscr{F}^+_{  \alpha \beta }=-\frac{i}{2}\partial_{\dot\rho( \alpha}  W_{ \beta)},\qquad
D_{ \rho}\mathscr{F}^+_{  \alpha \beta }=\frac{i}{2}\epsilon_{ \rho( \alpha}\partial_{ \beta)\dot\rho} \overline W^{ \dot\rho}.
\label{DF+}
\end{equation}

In Euclidean or Split signature, one can define the anti-self-dual version of SYM theory  \cite{Gilson:1985uj, Volovich}:
\begin{equation}
	\label{self_dual_constraint_F}
	*\mathscr{F}_{ab}\stackrel{!}{=}-i\mathscr{F}_{ab}, \qquad *\mathscr{F}_{ab}\equiv \frac12\epsilon_{abcd}\mathscr{F}^{cd}
\end{equation}
which implies, in the chiral notation, the conditions
\begin{equation}
	\label{selfdualconstraints}
		\mathscr{F}^-_{\dot\alpha\dot\beta}=0,\qquad
		\overline W_{\dot\alpha}=0.
\end{equation}

\section{Rheonomic Action for Anti-Self-Dual Super-Maxwell}
In the Abelian case, eqs. (\ref{FFWW}), (\ref{FDWFDW}) and (\ref{DDD}) become
\begin{equation}
	\mathscr{F}= 
	\frac12\bigg(\mathscr{F}^+_{\alpha \beta }  (V^2_+)^{\alpha\beta}
 	+
	{i}{W}_{ \alpha} (V\bar\psi)^{\alpha} \bigg)
\end{equation}	
with
\begin{equation}
	\label{DD}
	\mathscr{F}^+_{\alpha \beta }=
	-\frac{1}{4}{D}_{(\alpha} W_{\beta)},\qquad
	\overline{{D}}_{\dot\beta} W_{ \alpha}=0,\qquad
	{D}^{\alpha} W_{ \alpha}=0
\end{equation}
and eqs.  (\ref{DF-}), (\ref{DF+}) become
\begin{equation}
	\label{DFWDF}
	0=\frac{i}{2}\epsilon_{\dot\rho(\dot\alpha}\partial_{\dot\beta)\rho}  W^{ \rho},\qquad
	\overline D_{\dot\rho}\mathscr{F}^+_{  \alpha \beta }=-\frac{i}{2}\partial_{\dot\rho( \alpha}  W_{ \beta)},\qquad
	D_{ \rho}\mathscr{F}^+_{  \alpha \beta }=0.
\end{equation}
Applying $\overline D^{\dot\alpha} $  on  eq. (\ref{DD}) we have 
	\begin{equation}
		\label{diracequation}
		0= \overline D^{\dot\alpha} D^{\alpha} W_{\alpha} \stackrel{(\ref{DD})}{=}
		\{ \overline D^{\dot\alpha}, D^{\alpha} \}  W_{\alpha} =2i \partial^{\dot \alpha \alpha}  W_{\alpha}  \qquad \implies\qquad \partial^{\dot \alpha \alpha}  W_{\alpha} =0
	\end{equation}
	which is photino's Dirac equation. Similarly, one can obtain the Maxwell equations
	 \begin{equation}
		\label{MxA}
		\partial^{\dot\alpha \alpha} \mathscr{F}^+_{\alpha \beta} =0. 
	\end{equation}
	Since $D_{\rho} \mathscr{F}^+_{\alpha\beta} =0$ we have that   $\mathscr{F}^+_{\alpha\beta}$ is an \textit{antichiral} superfield.  Its first component is
	\begin{equation}
		\mathscr{F}^+_{\alpha\beta} =\mathscr{B}^+_{\alpha\beta}+ \mathcal{O}(\theta, \bar\theta).
	\end{equation} 	
 Similarly,  $\overline{{D}}_{\dot\beta} W_{ \alpha}=0$ means that $W_{ \alpha}$ is \textit{chiral}  and therefore:
	\begin{equation}
		W_{\alpha} =-i\lambda_{\alpha} +\theta_{\alpha} D + (\sigma^{ab}\theta)_{\alpha}F_{ab} +(\theta\theta)\partial_{\alpha\dot\beta}\bar\lambda ^{\dot\beta},
	\end{equation}
	where $F_{ab}$ is the usual gauge field strength. Taking into account the fact that Dirac and Maxwell's equations hold, the auxiliary field is zero $D=0$; moreover, being this an anti-self-dual theory, we have $\overline{W}^{\dot\alpha}=0$ and therefore also its first component is zero $\bar\lambda^{\dot\alpha}=0$. Finally, if we   define
	$
		H^+_{\alpha\beta} \equiv (\sigma^{ab})_{\alpha}^{\;\;\;\gamma}F_{ab}\epsilon_{\gamma\beta}   
	$
	we obtain
	$
		W_{\alpha} =-i\lambda_{\alpha}+ H^+_{\alpha\beta}\theta^{\beta}.
	$
	Note that $H^+_{\alpha\beta}$ is symmetric in its indices.
	Using $	D_{\alpha}= \partial_{\alpha}+i \bar \theta^{\dot\beta}\partial_{\alpha\dot\beta}$ we find that the first component of $\mathscr{F}^+_{\alpha \beta }$ is given by
	\begin{equation}
		\label{BH}
		\begin{aligned}
		\mathscr{F}^+_{\alpha \beta }=
		-\frac{1}{4}{D}_{(\alpha} W_{\beta)}=\mathscr{B}^+_{\alpha\beta}+ \mathcal{O}(\theta, \bar\theta)
		=
		-\frac{1}{4} H^+_{\alpha\beta}+ \mathcal{O}(\theta, \bar\theta)
		\end{aligned}
	\end{equation}
	which establishes a relation between the first component of $\mathscr{F}^+_{\alpha \beta }$ and the second component of $W_{\alpha}$.

Given this hypothesis, we can build the   {rheonomic} Lagrangian for anti-self-dual super Maxwell theory
 \begin{equation}
 	\label{MxH}
 	\begin{aligned}
 	\mathcal{L}^{(4|0)} &= \mathcal{F}^+_{\alpha\beta}(V_+^2)^{\alpha\beta}  \wedge \mathscr{F} - 
 	\frac{1}{12} \mathcal{F}^+_{\alpha\beta} \mathcal{F}_+^{\alpha\beta}  V^4  
 	+\\&\quad +
 	x (\mathcal{W}_{\alpha} \epsilon^{ \beta\alpha} \mathcal{W}_{\beta}) (\bar \psi V\wedge V \bar\psi) 
 	+ y \mathcal{F}^+_{\alpha\beta} (V_+^2)^{\alpha\beta} \wedge  \mathcal{W}_\sigma(V\bar\psi)^{\sigma}  
 	+ z  \mathscr{F} \wedge \mathcal{W}_\alpha (V\bar\psi)^{\alpha}
 	\end{aligned}   
 \end{equation}
which can be obtained following the standard procedure  \cite{cube}. In eq. (\ref{MxH})   $x,y,z$ are three constants to be fixed. It is crucial to highlight the difference between two types of $(2|0)$-superfields: 
\begin{align}
	\label{Fpiuself}
	\mathscr{F}= 
	\diff \mathscr{A},
	\qquad
	{\mathcal{F}}= 
	\frac12\bigg({\mathcal{F}}^+_{\alpha \beta }  (V^2_+)^{\alpha\beta}
	+
	{i}{\mathcal{W}}_{ \alpha} (V\bar\psi)^{\alpha} \bigg).
\end{align}	
The field strength  $ {\mathcal{F}}$, containing ${{\mathcal{F}}}^+_{\alpha\beta}$ and $ \mathcal{W}_{\alpha}$, is an \textit{auxiliary} anti-self-dual 2-form;  whereas  $\mathscr{F}$ is the   field strength of the \textit{physical}
gauge potential $ {\mathscr{A}}$, that is  $ \mathscr{F} = \diff {\mathscr{A}}$, and it contains  the \textit{physical} degrees of freedom. 
The relation between the auxiliary $ {\mathcal{F}}$ and the physical $ {\mathscr{F}}$ is given by the EL  equations in superspace: writing those for the auxiliary superfields $\mathcal{F}^+_{\alpha\beta}, \mathcal{W}^\alpha$ one obtains algebraic equations through which one can express them in terms of the physical ones; whereas writing the EL superspace equations for the physical  $ {\mathscr{A}}$ one obtains the dynamical equations of motion.  These are respectively given by:
\begin{align}
	&(V_+^2)^{\alpha\beta} \wedge \bigg(  \mathscr{F} + y  \mathcal{W}_\sigma (V\bar\psi)^{\sigma}  \bigg) 
	- \frac16 \mathcal{F}_+^{\alpha\beta} V^4 =0\label{prima_eq}
	\\&	 
	2 x \mathcal{W}_\alpha (\bar \psi V\wedge V \bar\psi)  - y  \mathcal{F}^+_{\gamma\beta} (V_+^2)^{\gamma\beta}\wedge (V\bar\psi)_\alpha 
	- z  \mathscr{F} \wedge (V\bar\psi)_\alpha =0 \label{second_eq}
	\\&
	\diff \left(\mathcal{F}^+_{\alpha\beta}(V_+^2)^{\alpha\beta} + z \mathcal{W}_\alpha (V\bar\psi)^\alpha  \right) =0\,. \label{eqmoto}
\end{align}

Inserting the general expression of $\mathscr{F}$   in eq. (\ref{prima_eq})  we obtain $
\overline{W}_{\dot\alpha} =0, 
\mathscr{F}_{\alpha\beta } =0, 
\mathscr{F}_{\alpha\dot{\beta} } =0, 
\mathscr{F}_{\dot{\alpha}\dot{\beta} } =0
$ and $\mathcal{F}_+^{\alpha\beta}= 6 \mathscr{F}_+^{\alpha\beta}, \mathcal{W}_{\alpha}=-\frac{i}{2y}W_{\alpha}$ 
(\textit{rheonomic parametrization}).  Notice that this equation does not give any information about $ \mathscr{F}^-_{\dot\alpha \dot\beta }  $ since $(V_+^2)^{ \alpha \beta} \wedge  (V_-^2)^{\dot\sigma\dot \tau}=0$. Choosing $z=2i$, eq. (\ref{second_eq}) fixes all coefficients $   y=-\frac{i}{6},   x=\frac16$ and  it gives the anti-self-duality condition\footnote{As a consequence,  projecting eq. (\ref{second_eq}) down to spacetime, we lost the anti-self-duality condition. } $
\mathscr{F}^-_{\dot\alpha \dot\beta }  =0$. Finally, eq. (\ref{eqmoto})  sets the theory on shell. Summarizing, eqs. (\ref{prima_eq}) (\ref{second_eq}) (\ref{eqmoto}) are intended on the entire superspace; they imply the on-shell conditions \eqref{diracequation}, \eqref{MxA} and the anti-self-duality condition. Nonetheless,   they are not derived from an action principle.

Using the rheonomic parametrizations, we get 
\begin{equation}
	\label{lagreo}
	\mathcal{L}^{(4|0)}=
	\frac{1}{12}\mathcal{F}^+_{\alpha\beta}	\mathcal{F}_+^{\alpha\beta} (V^4)
	+\frac{i}{6}\mathcal{F}^+_{\sigma\tau} (V_+^2)^{\sigma\tau}  \wedge \mathcal{W}_\alpha (V\bar\psi)^{\alpha} 
	+\frac13 \mathcal{W}_{\gamma} \epsilon^{ \delta \gamma} \mathcal{W}_{\delta}\epsilon_{\sigma\alpha} 
	( V \bar \psi )^{\sigma} \wedge (V  \bar\psi)^{\alpha}.
\end{equation}
One can easily show that the condition $\mathscr{F}^-_{\dot\alpha \dot\beta }  =0$   implies the Maxwell equation $\partial^{\dot\alpha \alpha} \mathscr{F}^+_{\alpha \beta} =0. $ As a consequence, the term $\mathcal{F}^+_{\alpha\beta}	\mathcal{F}_+^{\alpha\beta} $ is a total derivative\footnote{ This happens only in the abelian case. The non-abelian case is not a total derivative since the curvature has non-linear terms. } and the corresponding action is null. Projecting the Lagrangian down to spacetime, we get
\begin{eqnarray}
	\label{MxL}
	\mathcal{L}^{(4|0)}_{\mathrm{Spacetime}} = 
	3 \mathscr{B}^+_{\alpha\beta} \mathscr{B}_+^{\alpha\beta}  V^4
\end{eqnarray}
which is a total derivative, as mentioned. Therefore, we have two different problems. From one side, we cannot write the Lagrangian for the fermion nor its equation of motion. From the other side, the bosonic action is null unless there are some non-trivial boundary terms. Notice that also the interactions disappear due to the abelian nature of the present model, and a chiral spinor $\mathcal{W}_\alpha$ is not sufficient to build a vector current.

\section{Sen's Mechanism in components}
\label{Mechanism}

Sen's mechanism was applied to    $D= (4n + 2)$-dimensional theories \cite{Sen:2015uaa,Sen:2015nph,Sen:2019qit}  where the kinetic terms for self-dual field strength vanish. Here we adapt it to the cases $D=(4n+4)$ where the fermionic kinetic term is absent complementarily. To do so, let's introduce an \textit{auxiliary} field $\bar\chi$ in the spacetime Lagrangian (\ref{MxL})  with its kinetic term and an interaction term with the gaugino $\lambda$. If, following Sen's notation, $\mathcal{L}^{'}_{\mathrm{s.t.}}(\lambda,\Phi)$ is the Lagrangian of $\lambda_{\alpha}$ and any other field $\Phi$ of the theory, then we have 
\begin{equation}
	\label{SxA}
	{\mathcal L}_{ \mathrm{Sen}} = \frac12  \partial_{\alpha}^{\;\,\dot\alpha}\bar\chi_{\dot\alpha } \epsilon^{\alpha\beta}  \partial_{\beta}^{\;\,\dot\beta}\bar\chi_{\dot\beta }
	+ \partial^{\alpha\dot\alpha} \bar\chi_{\dot\alpha } \lambda _\alpha + 
	\mathcal{L}^{'}_{\mathrm{s.t.}}(\lambda,\Phi).
\end{equation}
The first two terms correspond to the  $\diff P^{(4)}\wedge * \diff P^{(4)} -\diff P^{(4)} \wedge Q^{(5)}$ of Sen’s Lagrangian \cite{Sen:2015nph}. The role of
the self-dual field $ Q^{(5)}$ is replaced here by the chiral gaugino $\lambda_{\alpha}$, and the role of the auxiliary field $ P^{(4)}$ is
played by $\bar\chi_{\dot\alpha}$. 

The equations of motion are 
\begin{equation}
	\label{firsteqmotion}
	\partial_{\alpha}^{\;\;\dot\alpha}\partial^{\alpha\dot\beta}\bar\chi_{\dot\beta}+\partial^{\alpha\dot\alpha}\lambda_{\alpha}=0,\qquad
	-\partial^{\alpha\dot\alpha} \bar\chi_{\dot\alpha }
	+
	\pdv{	 }{\lambda_{ \alpha}}\mathcal{L}^{'}_{\mathrm{s.t.}}(\lambda,\Phi)
	=0.
\end{equation}
Applying $\partial_{\alpha}^{\;\;\dot\beta}$ on the 
second equation and using the first one, we obtain
\begin{equation}
	\partial^{\alpha\dot\beta}\lambda_{\alpha}-\partial^{\alpha\dot\beta}\pdv{\mathcal {L}'_{\mathrm{s.t.}}}{\lambda^{ \alpha}}  =0
\end{equation}
which is the Dirac equation.
The last equation involves only the self-dual field $\lambda_{\alpha}$, and for any solution, we can use eq. \eqref{firsteqmotion} 
to compute $\bar\chi_{\dot\alpha }$. 
The crucial point is that, as well as for the Sen's mechanism in Type IIB supergravity,  the dofs of this extra
field
\footnote{
	Of course, at the level of spacetime action, in principle, one could introduce not only the $\bar\chi_{\dot\alpha}$ but also its supersymmetric partner, let's call it $k^-_{\dot\alpha\dot\beta}$ (the first component of the auxiliary superfield $\mathcal{K}^-_{\dot\alpha\dot\beta}$). However, this should be introduced without coupling it to the theory so that the Lagrangian would be trivially modified by an influent term $k_-^{\dot\alpha\dot\beta}k^-_{\dot\alpha\dot\beta}$. The upshot is that, in principle, the entire supermultiplet's fluctuations decouple from the theory. 
}
must decouple from the interacting sector, and therefore it will   not have physical
relevance. 
This decoupling can be seen at the level of equations of motion but not at the level
of the action.  
For a given $\lambda_{\alpha}$, two different solutions $\bar\chi_{\dot\alpha }$ and $\bar\chi'_{\dot\alpha }=\bar\chi_{\dot\alpha }+\Delta \bar\chi_{\dot\alpha }$ 
will differ by a free field $\Delta \bar\chi_{\dot\alpha }$ which satisfy 
the free second-order equation of motion 
\begin{equation}
	\label{chi_free}
	\partial_{\alpha}^{\;\;\dot\alpha}\partial^{\alpha\dot\beta} \Delta \bar\chi_{\dot\beta}=0.
\end{equation}
 It should be noted that this also 
matches the result of Sen: indeed, the kinetic term for $P_4$ in his Lagrangian has the wrong sign, leading to potential 
instabilities in the path integral. Nevertheless, since the field $\Delta P_4$ decouples from the theory, this amounts only 
to an overall regularization of the path integral. Conversely, for fermions, a second-order differential equation might lead to problems (unphysical modes which carry negative probabilities) for $\Delta \bar \chi_{\dot\alpha}$, but again they decouple from the theory.

The gaugino acquires an R-symmetry index $\lambda_{A\alpha}$ and for extended supersymmetry and eq. (\ref{SxA}) should be modified accordingly
\begin{equation}
	{\mathcal L}_{ \mathrm{Sen}} = \frac{1}{2} g^{AB}\partial_{\alpha}^{\;\,\dot\alpha}\bar\chi_{A\dot\alpha } \epsilon^{\alpha\beta}  \partial_{\beta}^{\;\,\dot\beta}\bar\chi_{B\dot\beta }
	+ \partial^{\alpha\dot\alpha} \bar\chi_{A\dot\alpha } \lambda^A _\alpha + 
	\mathcal{L}^{'}_{\mathrm{s.t.}}(\lambda,\Phi).
\end{equation}
The symmetric matrix $g^{AB}$ is  
a metric in the R-symmetry space. In $N=4$ case,   the $SU(4)$ 
R-symmetry is broken to $  SO(4)$. In   $N=2$, the $U(2)$ R-symmetry 
is broken to $SO(2) \times SO(2)$ as required for self-dual  super Maxwell multiplet. The symmetry of $g^{AB}$ is due to the fermionic 
nature of $\ chi$s and the contraction of the spinorial indices with $\epsilon$-tensors.

A similar strategy can be applied for the bosonic sector by  introducing two auxiliary fields: $\mathcal{G}_{\alpha\dot\beta}$ and $Q_+^{\alpha\beta}$. Notice the difference between $\mathscr{B}_+^{\alpha\beta}$ and $Q_+^{\alpha\beta}$: the Maxwell equation for the former follows from the Bianchi identity; the latter instead is the analog of $Q^{(5)}$ in Sen's paper, i.e. it is not a field strength (hence the Bianchi identity is not ensured) and it is a self-dual field. The resulting  Lagrangian is 
\begin{equation}\label{bosonic_lagr}
	\mathcal{L_{\mathrm{Sen, Q}}}=\frac12 \partial_{\alpha}^{\;\;\dot\alpha}\mathcal{G}_{\beta\dot\alpha}\,\epsilon^{\beta\rho}\,
	\partial^{\alpha\dot\gamma}\mathcal{G}_{\rho\dot\gamma}
	+
	\partial_{\alpha}^{\;\;\dot\alpha}\mathcal{G}_{\beta\dot\alpha}Q_+^{\alpha\beta}+\mathcal{L}'(Q,\Phi).
\end{equation}
The equations of motion are 
\begin{equation}
	\partial_{\beta}^{\;\;\dot\beta}
	\partial^{\beta\dot\alpha}
	\mathcal{G}^{\alpha}_{\;\;\dot\alpha}
	+
	\partial_{\beta}^{\;\;\dot\beta}Q_+^{\beta\alpha}=0,\qquad
	\partial^{\alpha\dot\rho}\mathcal{G}^{\beta}_{\dot\rho}+\pdv{\mathcal{L'}}{Q^+_{\alpha\beta}}=0
\end{equation}
and applying $\partial_{\alpha}^{\;\;\dot\gamma}$  to the second equation and using the first one, we obtain the Maxwell equation:
\begin{equation}
	\partial_{\alpha}^{\;\;\dot\beta}Q_+^{\alpha\beta}-\partial_{\alpha}^{\;\;\dot\beta}\pdv{\mathcal{L'}}{Q^+_{\alpha\beta}}=0.
\end{equation}
Note that, for a given $Q_+^{\alpha\beta}$, two solutions $\mathcal{G}_{\beta\dot\alpha}$ and $\mathcal{G}_{\beta\dot\alpha}'=\mathcal{G}_{\beta\dot\alpha}+\Delta \mathcal{G}_{\beta\dot\alpha}$ differ by a field $\Delta \mathcal{G}_{\beta\dot\alpha}$ which satisfies $
	\partial_{\beta}^{\;\;\dot\beta}
	\partial^{\beta\dot\alpha}\Delta \mathcal{G}^{\alpha}_{\;\;\dot\alpha}=0$, in analogy to the fermionic case.

\section{The Sen's Rheonomic action}\label{sen_rheon_action}

A meaningful action principle on a supermanifold ${\mathcal {SM}}^{(4|4)}$ is defined as an integral form Lagrangian ${\mathcal L}^{(4|4)}$. This is built in terms of the rheonomic  ${\mathcal L}^{(4|0)}$, see eq. (\ref{lagreo}), and in terms of an integral form $\mathbb{Y}^{(0|4)}$ (PCO) as follows
\begin{eqnarray}
	\label{AxA}
	{\mathcal L}^{(4|4)} = \mathcal{L}^{(4|0)} \wedge \mathbb{Y}^{(0|4)}.
\end{eqnarray}
$\mathbb{Y}^{(0|4)}$  is the Poincaré dual of the embedding of the spacetime $\mathcal{M}^{(4)}$ into the  supermanifold ${\mathcal {SM}}^{(4|4)}$ 	\cite{ Castellani:2015ata,	
	Castellani:2017fhi }. In the following, we consider two possible PCOs which differ by  $\diff$-exact terms:
\begin{eqnarray}
	\mathbb{Y}_{\mathrm{Standard}}^{(0|4)} &=&  (\theta\theta)(\bar\theta\bar\theta)\delta^4(\psi) \,, \label{PCOO} \\
	\mathbb{Y}_{\mathrm{Susy}}^{(0|4)} &=& 	\bigg(-4(\theta V\bar\iota)(\bar\theta V\iota)
	-(\theta\theta)(\iota \, V\wedge V\iota)
	+(\bar\theta\bar\theta)(\bar\iota\, V\wedge V \bar\iota)\bigg)\delta^4(\psi).\label{PCOsusy}
\end{eqnarray}
The  {operators} $\iota$, $\bar\iota$ read
\begin{equation}
	\label{iota}
	\iota_{\alpha}\equiv \iota_{D_{\alpha}}=\iota_{\partial_{\alpha}}+i\bar{\theta}^{\dot{\alpha}} \iota_{\alpha\dot{\alpha}},\qquad
	\bar{\iota}_{\dot{\alpha}}\equiv \iota_{\bar{D}_{\dot{\alpha}}}=\bar{\iota}_{\partial_{\dot{\alpha}}}+i{\theta}^{{\alpha}} \iota_{\alpha\dot{\alpha}}
\end{equation}
where $\iota_{\alpha\dot{\alpha}}\equiv {2\partial_{\alpha\dot{\alpha}}}$, $\iota_{\partial_{\alpha}}$ and $\bar{\iota}_{\partial_{\dot{\alpha}}}$ are the contraction operators along the cooridnate vector fields.
The contraction operator $\iota_{\alpha\dot{\alpha}}$ is an odd differential operator dual to $V^{\alpha\dot\alpha}$, while $\iota_{\alpha }$ and $\bar{\iota}_{ \dot{\alpha}}$ are even and dual to $\psi^\alpha$ and $\bar\psi^{\dot\alpha}$ . 

Using eq. (\ref{iota}) and the standard PCO,  eq. (\ref{AxA}) becomes 
\begin{equation}
	\label{lstandard}
	{\mathcal L}^{(4|4)}_{\mathrm{Standard}} =\mathcal{L}^{(4|0)}\wedge	\mathbb{Y}^{(0|4)}_{\mathrm{Standard}}=	3 \mathscr{B}^+_{\alpha\beta} \mathscr{B}_+^{\alpha\beta}\; (\theta\theta)(\bar\theta\bar\theta)\, V^4\wedge\delta^4(\psi)
\end{equation}
where the last two terms in (\ref{lagreo}) are canceled out due to the presence of the deltas. 
Using the supersymmetric PCO, we get
\begin{equation}
	\label{lsusy}
	{\mathcal L}^{(4|4)}_{\mathrm{Susy}}  = 
	\mathcal{L}^{(4|0)} \wedge\mathbb{Y}_{\mathrm{Susy}}^{(0|4)}=	
	16\mathcal{W}^\gamma \mathcal{W}_\gamma 
	(\bar\theta \bar\theta)
	V^4\wedge						
	\delta^4(\psi)
\end{equation}
where the antisymmetry of $V$'s and the integration by parts have been exploited.
\subsection{Fermionic Sector}
For  Sen's mechanism, we introduce the following \textit{auxiliary} field-strength 
\begin{equation}
	\label{RxA}
	\mathcal{K} \equiv \frac12\bigg(\mathcal{K}^+_{ \alpha\beta } (V_+^2)^{\alpha\beta} +\mathcal{K}^-_{ \dot\alpha\dot\beta } (V_-^2)^{\dot\alpha\dot\beta}  +i\overline{ \mathcal{Z}}_{\dot\alpha}(V \psi)^{\dot\alpha}+i \mathcal{Z}_{\alpha}(V\bar\psi)^{\alpha} \bigg)
\end{equation}
with the usual conditions derived from the Bianchi identities:
\begin{equation}
	\label{DDDZ}\small
	\mathcal{K}^+_{ \alpha\beta }=-\frac14D_{(\alpha}\mathcal{Z}_{\beta)}\qquad\mathcal{K}^-_{ \dot\alpha\dot\beta }=-\frac14\overline D_{(\dot\alpha}\overline{ \mathcal{Z}}_{\dot\beta)},\qquad
	\overline{{D}}_{\dot\beta}\mathcal{Z}_{ \alpha}=0,\qquad
	{D}_{\beta}\overline{\mathcal{Z}}_{\dot\alpha}=0,\qquad
	\overline{{D}}_{\dot\alpha}\overline{\mathcal{Z}}^{\dot\alpha}={D}^{\alpha} \mathcal{Z}_{ \alpha}
\end{equation}
where the superfield $\mathcal{Z}$ is defined in such a way its first component is the auxiliary field $\chi$ and similarly for  $\overline{\mathcal{Z}}$ and $\bar\chi$. Solving eqs. (\ref{DDDZ}) we have that $\mathcal{Z}_\alpha$ is a \textit{chiral} superfield:
\begin{equation}
	\mathcal{Z}_\alpha =-i\chi_{\alpha} +\theta_{\alpha} \mathcal{E} + (\sigma^{ab}\theta)_{\alpha}K_{ab} +\partial_{\alpha\dot\alpha} \bar\chi^{\dot\alpha} (\theta\theta).
\end{equation}
Without imposing any anti-self-duality condition, we do \textit{not} get the equations of motion. This means that $\bar\chi$ is not null, and the same holds for the auxiliary field $\mathcal{E}$, which is not zero off-shell. Defining $
	\mathcal{H}^+_{\alpha\beta}\equiv (\sigma^{ab})_{\alpha}^{\;\;\;\gamma}K_{ab}\epsilon_{\gamma\beta}$,
we can rewrite this equation in the following way
\begin{equation}
	\label{Z}
	\mathcal{Z}_\alpha = -i\chi_\alpha + ( \epsilon_{\alpha\beta} {\mathcal{E}}+\mathcal{H}^+_{\alpha\beta})\theta^\beta +  \partial_{\alpha\dot\alpha} \bar\chi^{\dot\alpha} (\theta\theta).
\end{equation}
Similarly, defining  $
	\mathcal{H}^-_{\dot\alpha\dot\beta}\equiv \epsilon_{\dot\alpha\dot\gamma}(\bar\sigma^{ab})_{\;\;\dot\beta}^{\dot\gamma}K_{ab} $, 
we have 
$\overline{\mathcal{Z}}$ is an \textit{antichiral} superfield and therefore
\begin{equation}
	\label{barZ}
	\overline{\mathcal{Z}}_{\dot\alpha} =i \bar\chi_{\dot\alpha} + ( \epsilon_{\dot\alpha\dot\beta} {\mathcal{E}}-\mathcal{H}^-_{\dot\alpha\dot\beta})\bar\theta^{\dot\beta} - \epsilon_{\dot\alpha\dot\beta}\partial^{\dot\beta\alpha}  \chi_{ \alpha} (\bar\theta\bar\theta).
\end{equation}
Let's consider the following Rheonomic action
\begin{equation}
\label{rheonomic_lagrangian_second_attempt}
\boxed{ \mathbb{S}_F\equiv 
		{\int}_{\mathcal{SM}^{(4|4)}}  	\bigg(\mathcal{L}^{(4|0)}+\mathscr{L}_F^{(4|0)}\bigg)\wedge	\mathbb{Y}^{(0|4)}
	  }		
\end{equation}  
\begin{equation}
	\label{RxD}
	\setlength{\fboxsep}{7pt}
	\setlength{\fboxrule}{0.5pt}
	\fbox{$\begin{aligned}
		\mathcal{S}_F&= \int_{\mathcal{SM}^{(4|4)}}\mathscr{L}_F^{(4|0)}\wedge \mathbb{Y}^{(0|4)} 
		=\\&=
		\frac13\int_{\mathcal{SM}^{(4|4)}}
		\left(\frac{i}{24}
		\mathcal{Z}_\alpha (V \bar\psi)^{\alpha} \wedge \mathcal{W}_\beta (V \bar\psi)^{\beta} 
		+ 
		\diff \overline{ \mathcal{Z}}_{\dot\alpha}\wedge (V^3)^{\alpha\dot\alpha} \mathcal{W}_\alpha  \right)  \wedge \mathbb{Y}^{(0|4)} 
		+\\&+\;\;\;
		\int_{\mathcal{SM}^{(4|4)}} \left( \diff \mathcal{Z}_{\alpha} \overline\aleph_{\dot\beta}\wedge    (V^3)^{\alpha\dot\beta}
		+\frac12 \overline\aleph_{\dot\alpha}\overline\aleph_{\dot\beta}\epsilon^{\dot\alpha\dot\beta}V^4  \right)\wedge \mathbb{Y}^{(0|4)}
		+\\&+\;\;\;
		\int_{\mathcal{SM}^{(4|4)}} \left( 
		\diff \overline{\mathcal{Z}}_{\dot\alpha}  \aleph_{ \beta}\wedge    (V^3)^{\beta \dot\alpha}
		+\frac12  \aleph_{ \alpha} \aleph_{ \beta}\epsilon^{ \alpha \beta}V^4 \right)\wedge \mathbb{Y}^{(0|4)}
		+\\&
		-\frac{i}{24}
		\int_{\mathcal{SM}^{(4|4)}} 
		\left(
		\diff \overline{\mathcal{Z}}_{\dot\alpha} \wedge  \mathcal{Z}_{\alpha}  \bar\psi^{\dot\alpha}\wedge V^{\alpha\dot\beta}\wedge \bar\psi^{\dot\delta}\epsilon_{\dot\delta\dot\beta}+ 
		\diff {\mathcal{Z}}_{\alpha} \wedge  \overline{\mathcal{Z}}_{\dot\alpha}  \psi^{\alpha}\wedge V^{\beta\dot\alpha}\wedge \psi^{\delta}\epsilon_{\delta\beta}
		\right) \wedge \mathbb{Y}^{(0|4)}.
	\end{aligned}$}
\end{equation} 
$\overline\aleph_{ \dot \alpha}$ and $ \aleph_{   \alpha}$ are two auxiliary superfields which will be fixed in terms of $\mathcal{Z}_{\alpha}$ and  $\overline{\mathcal{Z}}_{\dot\alpha}$ by the equations of motion.
We want to test this action against the two   PCOs introduced above.

\paragraph{PCO Standard}
Inserting Eq. (\ref{PCOO}) in \eqref{RxD} we  are left with
\begin{equation}
	\label{RxF}
	\begin{aligned}
	\mathcal{S}_F &= \frac13\int_{\mathcal{SM}^{(4|4)}} \left(
	\diff\overline{ \mathcal{Z}}_{\dot\alpha}\wedge (V^3)^{\alpha\dot\alpha} \mathcal{W}_\alpha  \right)  \wedge  (\theta\theta)(\bar\theta\bar\theta) \delta^4(\psi)
	+\\&+
	\int_{\mathcal{SM}^{(4|4)}} \left( \diff \mathcal{Z}_{\alpha} \overline\aleph_{\dot\beta}\wedge    (V^3)^{   \alpha \dot\beta}
	+\frac12 \overline\aleph_{\dot\alpha}\overline\aleph_{\dot\beta}\epsilon^{\dot\alpha\dot\beta}V^4 
	\right) 
	\wedge (\theta\theta)(\bar\theta\bar\theta) \delta^4(\psi)
	+\\&+
	\int_{\mathcal{SM}^{(4|4)}} \left(
	\diff \overline{\mathcal{Z}}_{\dot\alpha}  \aleph_{ \beta}\wedge    (V^3)^{\beta\dot\alpha}
	+\frac12  \aleph_{ \alpha} \aleph_{ \beta}\epsilon^{ \alpha \beta}V^4 \right) 
	\wedge (\theta\theta)(\bar\theta\bar\theta) \delta^4(\psi)
	.   
	\end{aligned}
\end{equation}
Using $\overline D_{\dot\alpha}\mathcal{Z}_{\beta}=0$ we have 
 \begin{equation}
 	\begin{aligned}
 		\diff \mathcal{Z}_\gamma=
 		\partial_{\alpha \dot \alpha} \mathcal{Z}_\gamma  V^{\alpha\dot\alpha}+ 
 		D_{\beta}\mathcal{Z}_{\gamma}
 		\psi^\beta.
 	\end{aligned}
 \end{equation}
Considering only the  $V$-part of $\diff \overline{ \mathcal{Z}}_{\dot\alpha}$ and  $\diff \mathcal{Z}_{\alpha}$  we obtain
\begin{eqnarray}
	\mathcal{S}_F&=& \frac13\int_{\mathcal{SM}^{(4|4)}} \left(
	\partial^{\alpha \dot \alpha} \overline{\mathcal{Z}}_{\dot\alpha}   \mathcal{W}_\alpha   \right)    (\theta\theta)(\bar\theta\bar\theta) V^4 \wedge\delta^4(\psi)  \nonumber \\
	&+&  \int_{\mathcal{SM}^{(4|4)}} 
	\left(
	\partial^{\alpha}_{\;\; \dot \gamma} \mathcal{Z}_\alpha   
	\overline\aleph^{\dot\gamma}      
	+\frac12 \overline\aleph_{\dot\alpha}\overline\aleph_{\dot\beta}\epsilon^{\dot\alpha\dot\beta}
	+
	\partial_{\gamma}^{\;\; \dot \alpha} \overline{\mathcal{Z}}_{\dot\alpha}   \aleph^{ \gamma}  
	+\frac12 \aleph_{ \alpha} \aleph_{ \beta}\epsilon^{ \alpha \beta}
	\right)   (\theta\theta)(\bar\theta\bar\theta) V^4\wedge \delta^4(\psi).   \label{azionePCO} 
\end{eqnarray}
Writing the EL equations for $\overline\aleph_{\dot\alpha}$, $\aleph_{\alpha}$, $\mathcal{Z}_{\alpha}$ and  $\overline{\mathcal{Z}}_{\dot\alpha}$ we obtain respectively
\begin{equation}
	\overline\aleph^{\dot\alpha}=-
	\partial^{\alpha \dot \alpha} \mathcal{Z}_\alpha,\qquad
	\aleph^{\alpha}=-
	\partial^{\alpha \dot \alpha}\overline{ \mathcal{Z}}_{\dot\alpha},\qquad
	\partial^{\alpha\dot\alpha}\overline\aleph_{ \dot \alpha}=0,\qquad
	\partial^{\alpha\dot\alpha}\aleph_{  \alpha}=0.
\end{equation}
Substituting these relations into the action (\ref{azionePCO}), then extracting the lowest component of the superfields
(due to the presence of $(\theta\theta)(\bar\theta\bar\theta) $)  and  finally adding the result  to (\ref{lstandard}), we obtain 
\begin{equation}
	\begin{aligned}
	\mathbb{S}_F&\equiv 
	\int_{\mathcal{SM}^{(4|4)}} \bigg( \mathcal{L}^{(4|0)}+\mathscr{L}^{(4|0)}_F\bigg)\wedge	\mathbb{Y}^{(0|4)}_{\mathrm{Standard}} 
	=\\&=
	\int_{\mathcal{M}^{(4)}_{\mathrm{red}}} (\mathrm{Total \; Derivative}) +
	\int_{\mathcal{M}^{(4)}_{\mathrm{red}}}\left(
	\frac{1}{2}
	\partial^{\;\,\dot\alpha}_\alpha \overline{ \chi}_{\dot\alpha}     
	\partial^{ \alpha\dot\beta} \overline{ \chi}_{\dot\beta}+
	\partial^{\alpha \dot \alpha} \overline{\chi}_{\dot\alpha}   \lambda_\alpha 
	+\frac{1}{2} 
	\partial^{\alpha}_{\;\, \dot \gamma} \chi_\alpha  \partial^{\beta\dot \gamma} \chi_\beta 
	 \right)     V^4  
	\end{aligned}
\end{equation}
which is exactly the spacetime Lagrangian (\ref{SxA}) reproducing  Sen's mechanism (here we have an extra term $\frac{1}{2} 
\partial^{\alpha}_{\;\, \dot \gamma} \chi_\alpha  \partial^{\beta\dot \gamma} \chi_\beta $ for the chiral part, but it is inessential because it is free, there is not an interaction term between $\chi$ and $\lambda$).

\paragraph{PCO Supersymmetric}

Let's consider the supersymmetric PCO (\ref{PCOsusy}). From the first line of eq. (\ref{RxD}), it survives   the first term:
$$
	\begin{aligned}
		& \quad \;\frac{i}{72}
		\int_{\mathcal{SM}^{(4|4)}}  
		\mathcal{Z}_\alpha (V \bar\psi)^{\alpha} \wedge \mathcal{W}_\beta (V \bar\psi)^{\beta} 
		(\bar\theta\bar\theta) (\bar\iota V\wedge V \bar\iota) 
		\wedge 
		\delta^4(\psi) 
		= \frac{i}{3}
		\int_{\mathcal{SM}^{(4|4)}}  
		\mathcal{Z}_{\alpha} \mathcal{W}^{\alpha} 	 
		(\bar\theta\bar\theta)
		V^4
		\wedge 
		\delta^4(\psi) 
	\end{aligned}
$$
and the Berezin integral gives
\begin{equation}
	\label{final1}
	\frac{i}{3}
	\int_{\mathcal{SM}^{(4|4)}}  
	\mathcal{Z}_{\alpha} \mathcal{W}^{\alpha} 	 
	(\bar\theta\bar\theta)
	V^4
	\wedge 
	\delta^4(\psi) =
	\int_{\mathcal{M}^{(4)}_{\mathrm{red}}}  
	\bigg( -\frac{i}{2}\mathcal{H}^+_{\alpha\beta}  H_+^{\alpha\beta}  
	+\partial^{\alpha\dot\alpha} \bar\chi_{\dot\alpha}  
	\lambda_{\alpha}  \bigg)V^4.
\end{equation}

The second and third lines of eq. (\ref{RxD}) vanish, while the first term in the fourth line gives
\begin{equation}
	\begin{aligned}
		&\quad-\frac{i}{24}\int_{\mathcal{SM}^{(4|4)}} \left(
		\diff \overline{\mathcal{Z}}_{\dot\alpha} \wedge  \mathcal{Z}_{\alpha}  \bar\psi^{\dot\alpha}\wedge V^{\alpha\dot\beta}\wedge \bar\psi^{\dot\delta}\epsilon_{\dot\delta\dot\beta}
		\right) \wedge \mathbb{Y}^{(0|4)}_\mathrm{Susy}
		=+\frac{i}{2}
		\int_{\mathcal{SM}^{(4|4)}} 
		\partial_{\alpha}^{\;\,\dot\alpha}\overline{\mathcal{Z}}_{\dot\alpha}\mathcal{Z}^{\alpha}  
		V^4\wedge	(\bar\theta\bar\theta)	\delta^4(\psi)
		\end{aligned}
\end{equation}
whose Berezin integral reads
\begin{equation}
	\label{final3_1}
	+\frac{i}{2}
	\int_{\mathcal{SM}^{(4|4)}} 
	\partial_{\alpha}^{\;\,\dot\alpha}\overline{\mathcal{Z}}_{\dot\alpha}\mathcal{Z}^{\alpha}  
	V^4\wedge	(\bar\theta\bar\theta)	\delta^4(\psi)=
	\int_{\mathcal{M}^{(4)}_{\mathrm{red}}} 
	\bigg(\frac{1}{2}\partial_{\alpha}^{\;\,\dot\alpha}\bar\chi_{\dot\alpha}\partial^{\alpha\dot\beta} \bar\chi_{\dot\beta} \bigg)
	V^4.	 	 
\end{equation}
The second term in the fourth line gives 
\begin{equation}
	\label{final3_2}
	-\frac{i}{2}
	\int_{\mathcal{SM}^{(4|4)}} 
	\partial_{\alpha}^{\;\,\dot\alpha}{\mathcal{Z}}^{\alpha}\overline{\mathcal{Z}}_{\dot\alpha}  
	V^4\wedge	(\theta\theta)	\delta^4(\psi)=
	\int_{\mathcal{M}^{(4)}_{\mathrm{red}}} 
	\bigg(\frac12 \partial^\alpha_{\;\;\dot\alpha}\chi_\alpha\partial^{\beta\dot\alpha}\chi_\beta \bigg)
	V^4
\end{equation}
after we extracted the $(\bar\theta\bar\theta)$ term in the last step.

Taking into account the results (\ref{lsusy}), (\ref{final1}),  (\ref{final3_1}) and  (\ref{final3_2}) we obtain the following superspace action
\begin{equation}
	\begin{aligned}
	\mathbb{S}_F&= 
	{\int}_{\mathcal{SM}^{(4|4)}}  	\bigg(\mathcal{L}^{(4|0)}+\mathscr{L}^{(4|0)}_F\bigg)\wedge	\mathbb{Y}^{(0|4)}_{\mathrm{Susy}}
	=\\&=
	{\int}_{\mathcal{SM}^{(4|4)}} 
	\bigg(
	+16\mathcal{W}^\gamma \mathcal{W}_\gamma 
	(\bar\theta \bar\theta)
	+\frac{i}{3}
	\mathcal{Z}_{\alpha} \mathcal{W}^{\alpha} 	 
	(\bar\theta\bar\theta)
	+\frac{i}{2}
	\partial_{\alpha}^{\;\,\dot\alpha}\overline{\mathcal{Z}}_{\dot\alpha}\mathcal{Z}^{\alpha}  
	(\bar\theta\bar\theta)
	-\frac{i}{2}
	\partial_{\alpha}^{\;\,\dot\alpha}{\mathcal{Z}}^{\alpha}\overline{\mathcal{Z}}_{\dot\alpha}  
	(\theta\theta)
	\bigg)
	V^4
	\wedge 
	\delta^4(\psi) 
	\end{aligned}
\end{equation}
and taking the Berezin integral
\begin{equation}
	\mathbb{S}_F= 
	\int_{\mathcal{M}^{(4)}_{\mathrm{red}}}  
	\bigg(
	1152\mathscr{B}^+_{\alpha\beta} \mathscr{B}_+^{\alpha\beta}
	-\frac{i}{2}\mathcal{H}^+_{\alpha\beta}  H_+^{\alpha\beta}  
	+\partial^{\alpha\dot\alpha} \bar\chi_{\dot\alpha}  
	\lambda_{\alpha}  
	+\frac{1}{2}\partial_{\alpha}^{\;\,\dot\alpha}\bar\chi_{\dot\alpha}\partial^{\alpha\dot\beta} \bar\chi_{\dot\beta}
	+\frac12 \partial^\alpha_{\;\;\dot\alpha}\chi_\alpha\partial^{\beta\dot\alpha}\chi_\beta
	\bigg)V^4.
\end{equation}
Using   $  H_+^{\alpha\beta} =-4\mathscr{B}_+^{\alpha\beta}$ and demanding   the auxiliary field-stength to satisfy $  \mathcal{H}_+^{\alpha\beta}\propto\mathscr{B}_+^{\alpha\beta}$  one obtains
\begin{equation}
	\mathbb{S}_F= 
	\int_{\mathcal{M}^{(4)}_{\mathrm{red}}} (\mathrm{Total \; Derivative}) +
	\int_{\mathcal{M}^{(4)}_{\mathrm{red}}}  
	\bigg(
	\partial^{\alpha\dot\alpha} \bar\chi_{\dot\alpha}  
	\lambda_{\alpha}  
	+\frac{1}{2}\partial_{\alpha}^{\;\,\dot\alpha}\bar\chi_{\dot\alpha}\partial^{\alpha\dot\beta} \bar\chi_{\dot\beta}
	+\frac12 \partial^\alpha_{\;\;\dot\alpha}\chi_\alpha\partial^{\beta\dot\alpha}\chi_\beta
	\bigg)V^4.
\end{equation}
 
\subsection{Bosonic Sector}
A similar idea can be implemented in the bosonic sector through the following auxiliary superfield
$$\small
	\begin{aligned}
		\mathcal{Y}_{\alpha\dot\beta}=&  \mathcal{G}_{\alpha\dot\beta} 
		+ 
		\Sigma_{\alpha\dot\beta  \gamma} \theta^{  \gamma}
		+ 
		\Gamma_{\alpha\dot\beta \dot\gamma}\bar \theta^{\dot \gamma}
		+
		M_{\alpha\dot\beta}( \theta \theta)
		+
		N_{\alpha\dot\beta}(\bar \theta\bar \theta)
		+\mathcal{G}_{\alpha\dot\beta\rho\dot\rho} \theta^{\rho}\bar\theta^{\dot\rho}
		+ 
		\bar\phi_{\alpha\dot\beta\dot\gamma}\bar\theta^{\dot\gamma}(\theta\theta)
		+
		\psi_{\alpha\dot\beta\gamma}\theta^{\gamma}(\bar\theta\bar\theta  )
		+
		\Phi_{\alpha\dot\beta}(\theta\theta)(\bar\theta\bar\theta).
	\end{aligned}
$$
By requiring  the condition $	D^{\beta}\mathcal{Y}_{  \alpha \dot\beta }=0 $
we obtain several constraints on the superfield's components. In particular, we have $ \Sigma^{\;\;\;\;\,\beta}_{\alpha\dot\beta }=0, M_{\alpha\dot\beta}=0, \mathcal{  G}^{\;\,\dot\eta}_{\alpha \;\,\sigma \dot\eta}  =+i \partial_{\sigma}^{\;\;\dot\beta} 
\mathcal{G}_{\alpha\dot\beta} $ and $\mathcal{Y}$ becomes:
\begin{equation}
	\begin{aligned}
		\mathcal{Y}_{\alpha\dot\beta}=&  \mathcal{G}_{\alpha\dot\beta}
		+ 
		\Gamma_{\alpha\dot\beta \dot\gamma}\bar \theta^{\dot \gamma}
		+
		N_{\alpha\dot\beta}(\bar \theta\bar \theta)
		+ \mathcal{G}_{\alpha\dot\beta\rho\dot\rho} \theta^{\rho}\bar\theta^{\dot\rho}
		+
		\bar\phi_{\alpha\dot\beta\dot\gamma}\bar\theta^{\dot\gamma}(\theta\theta)
		+
		\psi_{\alpha\dot\beta\gamma}\theta^{\gamma}(\bar\theta\bar\theta  )
		+
		\Phi_{\alpha\dot\beta}(\theta\theta)(\bar\theta\bar\theta).
	\end{aligned}
\end{equation}
Similarly, by introducing  $\mathcal{\overline Y}_{\alpha\dot\beta}$ and requiring
$\overline D^{\dot\alpha}\mathcal{\overline Y}_{  \alpha \dot\beta }=0$ we obtain 
$
\overline{\Sigma}^{\;\;\;\;\,\dot\alpha}_{\alpha\dot\beta }=0,
M^*_{\alpha\dot\beta}=0, 
\mathcal{  \overline G}^{\;\,\dot\eta}_{\alpha \;\,\sigma \dot\eta}  =-i \partial_{\sigma}^{\;\;\dot\beta} 
\mathcal{  \overline G}_{\alpha\dot\beta}$ and the superfield
\begin{equation}
	\begin{aligned}
		\mathcal{\overline Y}_{\alpha\dot\beta}=&  \mathcal{\overline G}_{\alpha\dot\beta} 
		+ 
		\overline \Gamma_{\alpha\dot\beta \gamma}  \theta^{  \gamma}
		+
		N^*_{\alpha\dot\beta}(  \theta  \theta)
		+\mathcal{\overline G}_{\alpha\dot\beta\rho\dot\rho} \theta^{\rho}\bar\theta^{\dot\rho}
		+
		\phi_{\alpha\dot\beta \gamma} \theta^{ \gamma}(\bar\theta\bar\theta)
		+
		\bar \psi_{\alpha\dot\beta\dot\gamma}\bar\theta^{\dot\gamma}(\theta\theta  )
		+
		\Phi^*_{\alpha\dot\beta}(\theta\theta)(\bar\theta\bar\theta).
	\end{aligned}
\end{equation}
Finally, by imposing the constraint ${D}^{\sigma}\overline D^{\dot\sigma}\mathcal{  Y}_{\alpha\dot\beta}=-\overline{D}^{\dot\sigma}D^{\sigma}\mathcal{ \overline Y}_{\alpha\dot\beta}$ we obtain $\mathcal{  \overline G}^{\;\,\dot\eta}_{\alpha \;\,\sigma \dot\eta}  =i \partial_{\sigma}^{\;\;\dot\beta} 
\mathcal{G}_{\alpha\dot\beta} $, that is $
\Rightarrow \mathcal{\overline G}_{\alpha\dot\beta}=- \mathcal{G}_{\alpha\dot\beta} +c_{\alpha\dot\beta} $.
We can now write the following rheonomic action for the bosonic sector
\begin{equation}
	\boxed{\begin{aligned}
			\mathcal{S}_B&=
			\int_{\mathcal{SM}^{(4|4)}}\mathscr{L}_B^{(4|0)} \wedge \mathbb{Y}^{(0|4)} 
			= \\&=
			\int_{\mathcal{SM}^{(4|4)}}
			\left(
			\frac{i}{12}\mathcal{\overline Y}_{\alpha\dot\beta}(V\bar\psi)^{\gamma}\wedge\Omega_{+\gamma}^{\alpha}(V\psi)^{\dot\beta}
			+
			\diff  { \mathcal{Y}}_{\alpha\dot\beta}\epsilon^{\beta\alpha}\wedge (V^3)^{\gamma\dot\beta} \Omega^+_{\beta\gamma}  \right)  \wedge \mathbb{Y}^{(0|4)} 
			\\&+
			\int_{\mathcal{SM}^{(4|4)}} \left( \diff  \mathcal{Y}_{\alpha\dot\beta}  \aleph_{\gamma\beta}\wedge    (V^3)^{\gamma\dot\beta}\epsilon^{\beta\alpha}
			+\frac12 \aleph_{\rho\gamma} \aleph_{\sigma\tau}\epsilon^{\sigma\rho}\epsilon^{\gamma\tau}V^4
			\right)\wedge \mathbb{Y}^{(0|4)}
			\\&
			+\int_{\mathcal{SM}^{(4|4)}} 
			\left(\frac{i  }{96}
			(\diff\mathcal{Y}_{\alpha\dot\alpha}   \overline{\mathcal{Y}}^{\alpha\dot\alpha} 
			- \mathcal{Y}_{\alpha\dot\alpha}   \diff\overline{\mathcal{Y}}^{\alpha\dot\alpha} 
			) \wedge 
			(\bar\psi V\psi)
			\right) 
			\wedge \mathbb{Y}^{(0|4)}.
	\end{aligned} }
\end{equation}
By using similar calculations, one can show that $\mathcal{S}_B$ gives the bosonic spacetime action \eqref{bosonic_lagr} for both the standard and the supersymmetric\footnote{Notice that, when considering $\mathbb{Y}_{\mathrm{Susy}}$, one has to deal also with the higher components of the bosonic auxiliary superfield ${\Omega}^+_{\alpha\beta}$ associeted with $Q^+_{\alpha\beta}$. By requiring $D_{ \rho}\Omega^+_{  \alpha \beta }=0$, it can be written as $
	\Omega^+_{\alpha\beta}=  Q^+_{\alpha\beta}
	+ 
	\mathscr{P}_{\alpha \beta \dot\gamma}\bar \theta^{\dot \gamma}
	+
	\mathscr{N}_{\alpha \beta}(\bar \theta\bar \theta)
	+ Q^+_{\alpha \beta\rho\dot\rho} \theta^{\rho}\bar\theta^{\dot\rho}
	+\mathcal{O}(\theta^3,\bar\theta^3)
	$ 	where $
	Q^{+\;\eta}_{\;\;\alpha\; \;\eta \dot\rho}=+ i \partial^{\beta}_{\;\;\dot\rho} Q^+_{\alpha \beta }$. 
	When extracting the $\theta\bar\theta$-terms from the first term and the last line of $\mathcal{S}_B$,  we respectively obtain a $\mathscr{P}\overline{\Gamma}$-term and a $\Gamma\overline{\Gamma}$-term which decouples entirely from the theory, in analogy to the  $\partial\chi \partial\chi$-term in the fermionic sector.  } PCO.

\section{Coupling  to gravitino}
{{ The rheonomic approach is a powerful framework for describing rigid supersymmetric field theories and supergravities. Indeed, by writing the rheonomic Lagrangian in terms of the vielbein $V^a$ and the gravitino $\psi^A$ superforms, we can straightforwardly consider the coupling with supergravity. However, verifying Sen's mechanism for the anti-self-dual Super Maxwell model even in the presence of supergravity is a highly non-trivial goal that goes beyond the aim of this paper. Instead, as an intermediate step, we are going to analyze Sen's mechanism by taking into account the interactions with a non-dynamical gravitino. A dictionary between the fields appearing in Type IIB supergravity and those appearing in our model will be established, thus providing a parallelism between the two theories. This analysis extends the results obtained in the previous sections slightly beyond the rigid case and provides the first step toward a full coupling with supergravity.}}
 
The gravitino 1-superforms are given by
\begin{equation}
	\begin{aligned}
	\psi^\alpha&=\psi^\alpha_{\beta\dot\beta}V^{\beta\dot\beta}
	+\psi^\alpha_{\beta}\diff\theta^{\beta}
	+\psi^\alpha_{\dot\beta}\diff\bar\theta^{\dot\beta}\\
	\bar\psi^{\dot\alpha}&=\bar\psi^{\dot\alpha}_{\beta\dot\beta}V^{\beta\dot\beta}
	+\bar\psi^{\dot\alpha}_{\beta}\diff\theta^{\beta}
	+\bar\psi^{\dot\alpha}_{\dot\beta}\diff\bar\theta^{\dot\beta}
	\end{aligned}
\end{equation}
where $\psi^\alpha_{\beta\dot\beta}$, $\bar\psi^{\dot\alpha}_{\beta\dot\beta}$ are the so-called \textit{left-right gravitino wavefunctions}, namely the physical degrees of freedom.  $\psi^\alpha_{\beta}$, $\bar\psi^{\dot\alpha}_{\beta}$ and $\psi^\alpha_{\dot\beta}$, $\bar\psi^{\dot\alpha}_{\dot\beta}$ are the geometrical data of superspace.
 
From the rheonomic Lagrangian (\ref{lagreo}), projecting on spacetime with the gravitino wavefunction $\bar\psi^{\dot\alpha}=\bar\psi^{\dot\alpha}_{\beta\dot\beta}V^{\beta\dot\beta}$ we get
$$
	\begin{aligned}
		\frac{i}{6}\mathcal{F}^+_{\alpha\beta}&\mathcal{W}_\gamma
		(V_+^2)^{\alpha\beta}  \wedge(V\bar\psi)^{\gamma}\bigg|_{s.t.}
		=  -6
		\mathscr{B}^{+\gamma\zeta}\lambda_\gamma \bar\psi^{\dot\delta}_{\zeta\dot\delta}V^4
	\end{aligned}
$$
and
$$
	\begin{aligned}
		\frac{1}{3}\mathcal{W}_{\eta} &\epsilon^{ \delta\eta} \mathcal{W}_{\delta} \epsilon_{\beta\alpha}
		( V \bar \psi )^{\beta} \wedge (V  \bar\psi)^{\alpha}\bigg|_{s.t.}
		=
		-6\lambda^{\delta}   \lambda_{\delta}\bar\psi^{\dot\lambda}_{\gamma\dot\gamma}\epsilon^{\gamma\zeta}\bar\psi^{\dot\kappa}_{\zeta\dot\zeta}
		[
		\delta^{\dot\gamma}_{\dot\kappa}\delta^{\dot\zeta}_{\dot\lambda}
		+
		\delta^{\dot\zeta}_{\dot\kappa}\delta^{\dot\gamma}_{\dot\lambda}
		]V^4.
	\end{aligned}
$$
If, for simplicity, we set 
\begin{equation}
	(\mathscr{B}^+\bar\psi)^\alpha\equiv	\mathscr{B}^{+\alpha\zeta}  \bar\psi^{\dot\delta}_{\zeta\dot\delta},\qquad
	(\bar\psi\bar\psi)\equiv  \bar\psi^{\dot\lambda}_{\gamma\dot\gamma}\epsilon^{\gamma\zeta}\bar\psi^{\dot\kappa}_{\zeta\dot\zeta}
	[
	\delta^{\dot\gamma}_{\dot\kappa}\delta^{\dot\zeta}_{\dot\lambda}
	+
	\delta^{\dot\zeta}_{\dot\kappa}\delta^{\dot\gamma}_{\dot\lambda}]
\end{equation}
the spacetime action $A= A_1+A_2$ is 
\begin{equation}
	\label{sd1}
	\begin{aligned}
		A_1
		= 
		\int_{\mathcal{M}^{(4)}_{\mathrm{red}}} 
		\bigg(
		i\lambda^{\alpha} \partial_{\alpha\dot\beta}\bar\lambda^{\dot\beta}+i\bar\lambda_{\dot\alpha}\partial^{\dot\alpha\beta}\lambda_{\beta}
		-
		6
		\lambda_\alpha	(\mathscr{B}^+\bar\psi)^\alpha
		-6\lambda^{\delta}   \lambda_{\delta}(\bar\psi\bar\psi) \bigg) V^4
	\end{aligned}
\end{equation}
\begin{equation}
	\label{sd2}
	A_2=
	\int_{\mathcal{M}^{(4)}_{\mathrm{red}}} 
	\bigg(
	3 \mathscr{B}^+_{\alpha\beta} \mathscr{B}_+^{\alpha\beta}
	\bigg)V^4=	\int_{\mathcal{M}^{(4)}_{\mathrm{red}}} (\mathrm{Total \; Derivative}).
\end{equation}
Note that the Dirac term,  which vanishes for the anti-self-dual model, has been added to ease the comparison with   Sen's work \cite{Sen:2015nph}. Indeed we can establish a dictionary as in the following table.
\begin{table}[h]
	\begin{center}
		\begin{tabular}{cc}
			\hline
			\hline
			\textsc{  Type IIB SUGRA} & \textsc{  Anti-Self-Dual SYM} \\ 
			\hline
			\hline
			$ \hat F^{(5)}=F^{(5)}+B^{(2)}\wedge F^{(3)} $ & $\lambda_\alpha$\\
			$(B^{(2)}, F^{(3)} )\equiv M$  & $( \mathscr{B}^+_{\alpha\beta}, \bar\psi^{\dot\delta}_{\alpha\dot\beta}  )\equiv\Phi$ \\
			\hline
			\hline
		\end{tabular}
	\end{center}
\end{table}

\noindent We firstly notice that as ${S}_2$ in \cite{Sen:2015nph}  depends on  all   the   fields   except  $ C^{(4)}$, in our case $	A_2 $ depends on all  the  fields except $ \lambda_{\alpha}$.
The comparison with \cite{Sen:2015nph} reads
\begin{align}
	-\frac{1}{2}\int \hat{ F}^{(5)}\wedge * \hat{ F}^{(5)} 
	\qquad &\longleftrightarrow \qquad
	\int\bigg(i\lambda^{\alpha} \partial_{\alpha\dot\beta}\bar\lambda^{\dot\beta}+i\bar\lambda_{\dot\alpha}\partial^{\dot\alpha\beta}\lambda_{\beta}\bigg)\nonumber
\\
	\int F^{(5)}\wedge B^{(2)}\wedge F^{(3)}
	\qquad &\longleftrightarrow \qquad
	\int\bigg(
	-	6
	\lambda_\alpha	(\mathscr{B}^+\bar\psi)^\alpha
	-6\lambda^{\delta}   \lambda_{\delta}(\bar\psi\bar\psi)
	\bigg)\nonumber
\\
	{S}_2
	\qquad &\longleftrightarrow \qquad
	\int(\mathrm{Total \; Derivative}).\nonumber
\end{align}
In Type IIB, imposing  the    self-duality condition $* \hat{ F}^{(5)}=\hat{ F}^{(5)}$,   the kinetic term of $\hat{ F}^{(5)}$ vanishes identically.  In our case, $\bar\lambda_{\dot\alpha}=0$ implies the vanishing of the Dirac Lagrangian. 
The equations of motion for $\bar\lambda_{\dot\alpha}$ and $ \lambda_{ \alpha}$ are respectively given by  
\begin{equation}
	\label{bianchilambda}
	\partial^{\dot\alpha\alpha}\lambda_\alpha=0,\qquad
	2i
	\partial^{\dot\alpha\alpha} \bar\lambda_{\dot\alpha}
	-
	6
	(\mathscr{B}^{+ }  \bar\psi)^{\alpha} 
	+12\lambda^{\alpha} (\bar\psi\bar\psi)
	=0.
\end{equation}
The second equation parallels 
\begin{equation}
	\diff(*\hat{ F}^{(5)}- B^{(2)}\wedge F^{(3)})=0
\end{equation}
of Type IIB supergravity.
The equations of motion concerning the rest of the fields are
\begin{equation}
	0=\delta_{\Phi} A=\delta_{\Phi} A_1+\delta_{\Phi} A_2
\end{equation}
where $\delta_{\Phi}$ denotes the variation concerning all other fields collectively denoted by $\Phi$ at \textit{fixed} $\lambda$.

We use the correspondence listed in the following table for Sen's mechanism.
\begin{table}[h]
	\begin{center}
		\begin{tabular}{ccc}
			\hline
			\hline
			Superstring Field Theory&\textsc{  Type IIB SUGRA} & \textsc{  Anti-self-dual SM} \\ 
			\hline
			\hline
			$\tilde\psi$&$ P^{(4)} $ & $\bar\chi_{\dot\alpha}$\\
			\multirow{2}{*}{$\psi$}&$Q^{(5)} $ & $\lambda'_{\alpha}$\\
			&$(B^{(2)}, F^{(3)})\equiv M$  &$( \mathscr{B}^+_{\alpha\beta}, \bar\psi^{\dot\delta}_{\alpha\dot\beta}  )\equiv\Phi$ \\
			\hline
			\hline
		\end{tabular}
	\end{center}
\end{table}

\noindent We can write the following action
\begin{equation}
	\label{mathdsS'}
	A'= A'_1+A_2
\end{equation}
where
\begin{equation}
	\label{s'1sd}
	A'_1 = \int_{\mathcal{M}^{(4)}_{\mathrm{red}}} 
	\bigg(
	\frac12  \partial_{\alpha}^{\;\,\dot\alpha}\bar\chi_{\dot\alpha} \partial^{\alpha\dot\beta}\bar\chi_{\dot\beta}
	+
	\partial^{\alpha\dot\alpha} \bar\chi_{\dot\alpha} \lambda'_\alpha 
	+
	\mathcal {L}'_{\mathrm{s.t.}}(\lambda', \Phi)
	\bigg)V^4
\end{equation} 
and $A_2$ is the same in eq. (\ref{sd2}). 
$\mathcal {L}'_{\mathrm{s.t.}}(\lambda', \Phi)$ has to be determined by demanding that the equations of motion derived from
this action agrees with those derived in the previous sections.
The comparison between the action (\ref{s'1sd}) and \cite{Sen:2015nph} is now evident:
\begin{align}
	\frac12\int\diff P^{(4)}\wedge * \diff P^{(4)}
	\qquad &\longleftrightarrow \qquad
	\frac12\int  \partial_{\alpha}^{\;\,\dot\alpha}\bar\chi_{\dot\alpha} \partial^{\alpha\dot\beta}\bar\chi_{\dot\beta}\nonumber
	\\
	-\int \diff  P^{(4)}\wedge Q^{(5)}
	\qquad &\longleftrightarrow \qquad
	\int\partial^{\alpha\dot\alpha} \bar\chi_{\dot\alpha} \lambda'_\alpha\nonumber
	\\
	-\int B^{(2)}\wedge  F^{(3)}\wedge  Q^{(5)}+ \frac12 \int* (B^{(2)}\wedge F^{(3)})\wedge (B^{(2)}\wedge F^{(3)})
	\qquad &\longleftrightarrow \qquad
	\hat{A}'_1\equiv\int{\mathcal L}'_{\mathrm{s.t.}}(\lambda', \Phi)\nonumber
\end{align}	 
and similarly for the equations of motion 
\begin{equation}
	\label{SxB}
	\begin{aligned}
		\diff(*\diff P^{(4)}- Q^{(5)})&=0
		\qquad \longleftrightarrow \qquad 
		\partial_{\alpha}^{\;\;\dot\alpha}\partial^{\alpha\dot\beta}\bar\chi_{\dot\beta}+\partial^{\alpha\dot\alpha}\lambda'_{\alpha}=0
		\\
		\diff P^{(4)}+B^{(2)}\wedge F^{(3)}-*(\diff P^{(4)}+B^{(2)}\wedge F^{(3)})&=0
		\qquad \longleftrightarrow \qquad 
		-\partial^{\alpha\dot\alpha} \bar\chi_{\dot\alpha} + \frac{\partial  {\mathcal L}'_{\mathrm{s.t.}}}{\partial \lambda'_\alpha} =0\\
		\diff Q^{(5)}-\diff(B^{(2)}\wedge F^{(3)}) +\diff * (B^{(2)}\wedge F^{(3)})&=0
		\qquad \longleftrightarrow \qquad 
		\partial^{\alpha\dot\alpha} \lambda'_{\alpha} - \partial^{\alpha\dot\alpha} \frac{\partial  {\mathcal L}'_{\mathrm{s.t.}}}{\partial \lambda'^\alpha}=0. 
	\end{aligned}
\end{equation}

At the level of the action, the equivalence between $A'=A'_1+A_2$   and $A=A_1+A_2$ can be achieved by  
a suitable identification of
the field $\lambda'_\alpha$ with a combination of $\lambda_\alpha$ and $\mathscr{B}^+_{\alpha\beta}, \bar\psi^{\dot\delta}_{\alpha\dot\beta} $.  It follows 
\begin{enumerate}
	\item $\lambda_{\alpha} $, written in terms of $\lambda'_\alpha$,   should satisfy the equation of motion (\ref{bianchilambda}) as a consequence of the third equation of (\ref{SxB}).
	\item We must have $ \delta_{\Phi}A'_1=\delta_{ \Phi}A_1$.	
	\item By the above identification, a given solution to the equations of motion  of the action $A=A_1+A_2$ must give a set of solutions to those derived by
	$A'=A'_1+A_2$ which differ from each other by plane waves. The latter are free non-interacting degrees of freedom.
\end{enumerate}
Let's start analyzing the three points.

\begin{enumerate}
	
	\item 
	We consider the following equation
	\begin{equation}
		\label{lambdaR}
		0=-\partial_{\alpha}^{\;\;\dot\alpha} \lambda'^{\alpha} + \partial_{\alpha}^{\;\;\dot\alpha} \frac{\partial  {\mathcal L}'_{\mathrm{s.t.}}}{\partial \lambda'_\alpha}\equiv
		\partial_{\alpha}^{\;\;\dot\alpha} (-\lambda'^{\alpha} +R^{\alpha})
	\end{equation}
	where $R_{\alpha}$ is the analog of $R^{(5)}$  \cite{Sen:2015nph}. In other words $R_{\alpha}$ is defined by
	\begin{equation}
		\label{hatA}
		\delta \hat{A}'_1 =\delta  \int_{\mathcal{M}^{(4)}_{\mathrm{red}}}{\mathcal L}'_{\mathrm{s.t.}}(\lambda',\Phi) =
		\int_{\mathcal{M}^{(4)}_{\mathrm{red}}}
		\frac{\partial  {\mathcal L}'_{\mathrm{s.t.}}}{\partial \lambda'_\alpha} \delta\lambda'_{\alpha}+\delta_{\Phi}\hat{A}'_1 
		\equiv \int_{\mathcal{M}^{(4)}_{\mathrm{red}}} R^{\alpha}\delta\lambda'_{\alpha}+\delta_{\Phi}\hat{A}'_1.
	\end{equation}
	Following the procedure described by Sen, we should compare
	eq. (\ref{lambdaR}) and the second one of  (\ref{bianchilambda}) with $\bar\lambda_{\dot\alpha}=0$. In this way, we would obtain the above identification. However, this is not so straightforward. Indeed, in Type IIB supergravity, Sen uses a crucial numerical factor $1/2$ to ensure the matching. In our model, this can also be achieved using a non-trivial $(\bar\psi\bar\psi)$ term, as in the following equation:
	\begin{equation}
		\label{identification}
		\big(1+12  (\bar\psi\bar\psi)\big)(-\lambda'^{\alpha} +R^{\alpha})=
		-
		6
		(\mathscr{B}^+  \bar\psi)^{\alpha} 
		+12\lambda^{\alpha} (\bar\psi\bar\psi)
	\end{equation}	
	which gives two equations
	\begin{equation}
		-\lambda'^{\alpha} +R^{\alpha}=\lambda^\alpha,\qquad
		-\lambda'^{\alpha} +R^{\alpha}=-
		6
		(\mathscr{B}^+  \bar\psi)^{\alpha} 
	\end{equation}
	that is
	\begin{equation}
		\label{lambda_B_psi}
		\lambda^\alpha=
		-
		6
		(\mathscr{B}^+  \bar\psi)^{\alpha}  \quad \Rightarrow \quad
		R^{\alpha}=-
		6
		(\mathscr{B}^+  \bar\psi)^{\alpha}+\lambda'^{\alpha} .
	\end{equation}
	Notice that, in eq. (\ref{identification})  we are ignoring the dimensional mismatching. Indeed, recalling that 
	the mass dimensions are $[\bar\psi]=+3/2$ and  $[\lambda]=+3/2$,  eq. (\ref{identification}) should be modified, rescaling the gravitino terms by a suitable power of the Planck mass $m_{pl}=1/\sqrt{G_N}$.\footnote{Another possibility that mimics Sen's conventions is to assign dimension $+1$ to $\bar\chi$ and $+2$ to $\lambda$. This way, we cannot introduce a dimensionful mass parameter in the present equations.}
	
	\item  
	Let's find $\hat{A}'_1 $. From eq. (\ref{hatA}) we have
	\begin{equation}
		\delta \hat{A}'_1(\lambda',\Phi) = 
		\int_{\mathcal{M}^{(4)}_{\mathrm{red}}} 	\lambda'^{\alpha}\delta\lambda'_{\alpha}
		- \int_{\mathcal{M}^{(4)}_{\mathrm{red}}} 6(\mathscr{B}^+\bar\psi)^\alpha \delta\lambda'_{\alpha}
		+
		\delta_{\Phi}\hat{A}'_1(\lambda',\Phi).
	\end{equation}
	Using  $ \pdv{(\lambda'^{\rho}\lambda'_{\rho})}{\lambda'_{\alpha}}=-2\lambda'^\alpha$, we have
	\begin{equation}
		\label{hats'1}
		\hat{A}'_1 (\lambda',\Phi)= 
		\int_{\mathcal{M}^{(4)}_{\mathrm{red}}}	-\frac12\lambda'^{\alpha}\lambda'_{\alpha} 
		+ \int_{\mathcal{M}^{(4)}_{\mathrm{red}}}6(\mathscr{B}^+\bar\psi)^\alpha \lambda'_{\alpha}
		+
		\tilde{A}'_1(\lambda',\Phi).
	\end{equation}
	Now in order to find $	\tilde{A}'_1(\lambda',\Phi)$ we have to impose 
	$
	\delta_{\Phi}A'_1(\lambda',\Phi)=\delta_{ \Phi}A_1(\lambda,\Phi),
	$
	where
	\begin{equation}
		\label{vars'1}
		\begin{aligned}
			\delta_{\Phi}A'_1=\delta_{ \Phi}\hat{A}'_1
			&=
			\int_{\mathcal{M}^{(4)}_{\mathrm{red}}}  6\delta_{\Phi}(\mathscr{B}^+\bar\psi)^\alpha \lambda'_{\alpha}
			+
			\delta_{\Phi}\tilde{A}'_1(\lambda',\Phi)
		\end{aligned}
	\end{equation}
	and
	\begin{equation}
		\label{vars1}
		\delta_{\Phi}A_1=
		\int_{\mathcal{M}^{(4)}_{\mathrm{red}}} 
		\bigg(
		6\delta_{\Phi}	(\mathscr{B}^+\bar\psi)^\alpha\lambda_{\alpha}
		-6 \lambda^{\alpha}\lambda_{\alpha}\delta_{\Phi}(\bar\psi\bar\psi)
		\bigg).
	\end{equation}
	Notice that the variation $\delta_{\Phi}$ is performed only concerning the gauge field since we consider a \textit{non-dynamical} gravitino. Therefore $\delta_{\Phi}(\bar\psi\bar\psi)=0$. Using 
	this fact and the first equation of (\ref{lambda_B_psi}), we have
	\begin{equation}
		\delta_{\Phi}\tilde{A}'_1(\lambda',\Phi)=	-\int_{\mathcal{M}^{(4)}_{\mathrm{red}}}  6\delta_{\Phi}(\mathscr{B}^+\bar\psi)^\alpha \lambda'_{\alpha}
		-\int_{\mathcal{M}^{(4)}_{\mathrm{red}}} 
		36\delta_{\Phi}	(\mathscr{B}^+\bar\psi)^\alpha(\mathscr{B}^+\bar\psi)_{\alpha}
	\end{equation}
	from which
	\begin{equation}
		\begin{aligned}
			\tilde{A}'_1(\lambda',\Phi)&=-18	(\mathscr{B}^+\bar\psi)^\alpha(\mathscr{B}^+\bar\psi)_{\alpha}-6(\mathscr{B}^+\bar\psi)^\alpha\lambda'_{\alpha}.
		\end{aligned}
	\end{equation}
	Therefore,  eq. (\ref{hats'1}) becomes
	\begin{equation}
		\hat{A}'_1 (\lambda',\Phi)= 
		\int_{\mathcal{M}^{(4)}_{\mathrm{red}}} 
		\bigg\{-\frac12\lambda'^\alpha\lambda'_\alpha
		-18	(\mathscr{B}^+\bar\psi)^\alpha(\mathscr{B}^+\bar\psi)_\alpha
		\bigg\}.
	\end{equation}
	Finally,  we can substitute this equation in (\ref{s'1sd}), obtaining
	$$
	\boxed{
		A'_1 = \int_{\mathcal{M}^{(4)}_{\mathrm{red}}} 
		\bigg\{
		\frac12  \partial_{\alpha}^{\;\,\dot\alpha}\bar\chi_{\dot\alpha} \partial^{\alpha\dot\beta}\bar\chi_{\dot\beta}
		+
		\partial^{\alpha\dot\alpha} \bar\chi_{\dot\alpha} \lambda'_\alpha
		-\frac12\lambda'^\alpha\lambda'_\alpha
		-18(\mathscr{B}^+\bar\psi)^\alpha(\mathscr{B}^+\bar\psi)_\alpha
		\bigg\}V^4.
	}  
	$$

	\item The third point follows as explained in \cref{Mechanism} (see eq. (\ref{chi_free})).

\end{enumerate}
The action $A'_1$ correctly depends on the auxiliary fields $\bar\chi, \lambda'$ and on all the other fields of the theory, namely the gauge and gravitino fields. Moreover, the result obtained matches the result of Sen \cite{Sen:2015nph}: a     mass term for the chiral gaugino appears for consistency, and it parallels the term $\frac{1}{16}\int {Q^{(5)}}^T\mathcal{M}Q^{(5)}$ in Sen's action, while the squared term $(\mathscr{B}^+\bar\psi)^{\alpha}(\mathscr{B}^+\bar\psi)_{\alpha}$ parallels $\frac{1}{4}\int Y^{T}\mathcal{M}Y$.  However, 
the four-fermions term is missing, and we do not have a parallel for the term 
$\frac{1}{2}\int {Q^{(5)}}^{T} [\frac{1}{2}\mathcal{M}Y-(\zeta -\epsilon)Y]-\frac{1}{2}\int Y^{T} \zeta Y   $. This is 
probably due to the non-dynamical character of the gravitino considered here. Therefore, this point needs further clarification, which will undoubtedly be pursued in future works involving a complete coupling with supergravity.

\newpage
\section{Acknowledgment}

We want to thank L. Andrianopoli, L. Castellani, C.A. Cremonini, R. D'Auria, R. Noris, L.Ravera, and M. Trigiante for their helpful discussions. G.B. will like to thank INFN for the invitation to Alessandria. The work is partially supported by 
UPO research funds.

\vfill\eject
\end{document}